\documentclass[twocolumn,aps,prl,showpacs,superscriptaddress,notitlepage]{revtex4-1}
\usepackage{bbm}
\usepackage[babel]{csquotes}
\usepackage{braket}
\usepackage{graphicx}
\usepackage{amsmath, amsfonts, amssymb, graphics, setspace}
\usepackage[colorlinks=true,citecolor=blue,urlcolor=blue]{hyperref}
\usepackage{bbold}
\usepackage{braket}
\usepackage{amsthm}
\usepackage{subfigure}
\usepackage{mathrsfs}
\usepackage[usenames, dvipsnames]{xcolor}
\usepackage[a4paper,top=3cm,bottom=3cm,left=1.3cm,right=1.3cm]{geometry}

\usepackage{commath}

\begin{document}

\title{
Twin GHZ-states behave differently
}

\author{Gonzalo Carvacho}
\affiliation{Dipartimento di Fisica, ``Sapienza''
 Universit\`{a} di Roma, I-00185 Roma, Italy}
 
 \author{Francesco Graffitti}
\affiliation{Dipartimento di Fisica, ``Sapienza''
 Universit\`{a} di Roma, I-00185 Roma, Italy}
 
\author{Vincenzo D'Ambrosio}
\affiliation{Dipartimento di Fisica, ``Sapienza''
 Universit\`{a} di Roma, I-00185 Roma, Italy}
 
\author{Beatrix C. Hiesmayr}
\email{beatrix.hiesmayr@univie.ac.at}
\affiliation{University of Vienna, Faculty of Physics, Boltzmanngasse 5, 1090 Vienna, Austria}

\author{Fabio Sciarrino}
\email{fabio.sciarrino@uniroma1.it}
\affiliation{Dipartimento di Fisica, ``Sapienza''
 Universit\`{a} di Roma, I-00185 Roma, Italy}

\begin{abstract}
Greenberger-Horne-Zeilinger (GHZ) states and their mixtures exhibit fascinating properties. A complete basis of GHZ-states can be constructed by properly choosing local basis rotations. We demonstrate this experimentally for the Hilbert space $\mathcal{C}_2^{\otimes 4}$ by entangling two photons in polarisation and orbital angular momentum. Mixing GHZ-states unmasks different entanglement features based on their particular local geometrical connectedness. In particular, a specific GHZ-state in a complete orthonormal basis has a ``\textit{twin}'' GHZ-state for which equally mixing leads to full separability in opposition to any other basis-state.  Exploiting these local geometrical relations provides a toolbox for generating specific types of multipartite entanglement, each providing different benefits in outperforming classical devices. Our experiment, based on hybrid entangled entanglement, investigates these GHZ's properties showing a good stability and fidelity and allowing a scaling in degrees of freedom and advanced operational manipulations.
\end{abstract}

\maketitle

{\bf Introduction.}
Entanglement, a fundamental concept of quantum theory, can occur if states have to be described by tensored Hilbert spaces~\cite{Reventanglem}. Surprisingly, entanglement is not limited to physical distinguishable particles but exhibits itself also between different degrees of freedom~\cite{pub112,carlos,Pan,belldof,nagasanta,kwiatbarr,aielloleuchs}. Mathematically speaking, a physical system can be separable or entangled with respect to a chosen factorization of the total algebra which describes the quantum state. Indeed, the choice of factorization allows for pure states to switch unitary between separability and entanglement, however, usually the experimental setup fixes the factorization and applying local unitaries does not change the entanglement properties.

Genuine multipartite entangled states are of special interest since they are the extreme version of entanglement, that is all subsystems contribute to the shared entanglement feature~\cite{Reventanglem, detchall, Guhne20091}. Here, further coarse graining applies due to distinct physical properties of genuine multipartite entanglement, the Greenberger-Horne-Zeilinger (GHZ) states, the graph states, the $W$-states or Dicke-states are such examples.

Entanglement is nowadays a keystone in applied sciences, with applications ranging from quantum teleportation \cite{Bennet} to quantum dense coding \cite{dense},
quantum computation \cite{shor1995, steane1996, feynman1982} and quantum cryptography \cite{ekert1991,QKDE}.

Here we have produced two physical photons for which we consider the polarisation degree of freedom and a two-dimensional subspace of the orbital angular momentum (OAM) degree of freedom for each photon~\cite{Car12,sing,vazirioam,Nagali10,marrpadg,Torres,cabello,VVB}. Thus we explore a $16$ dimensional Hilbert space with the structure $\mathcal{C}_2\otimes\mathcal{C}_2\otimes\mathcal{C}_2\otimes\mathcal{C}_2$. Further classifications of the correlations between subsystems that may or may not be stronger than any correlations based on classical communications is needed in tackling specific quantum properties allowing to outperform classical algorithms. In our case one has three different ``depth'' of entanglement, the state can be tri-separable, bi-separable or genuine multipartite entangled between the different subsystems. This concept of $k$-separability is rather simple, however, due to its non-constructive nature, the problem of detecting it for mixed states is still open~\cite{witnessblatt,beatrix2}.

\vspace{0.2cm}

{\bf Multipartite Entanglement.}
\begin{figure}
\centering
\includegraphics[width=0.49\textwidth,keepaspectratio=true]{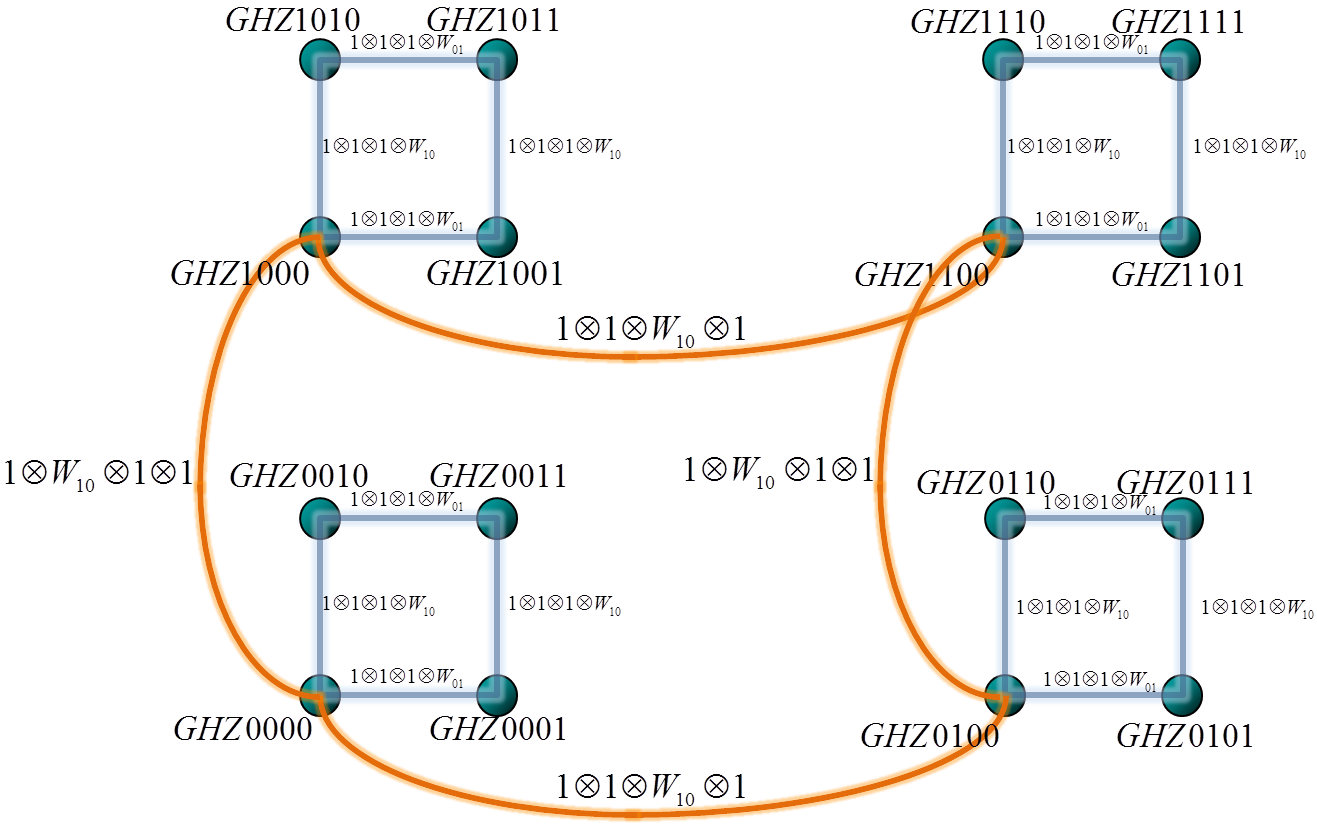}
\caption{GHZ basis geometry. By applying the Weyl operators $W_{k,l}$ (Pauli's operators) to the fourth subsystem it is possible to reach each quadrant's vertex.
In order to move horizontally (vertically) from one quadrant to another one it is necessary to apply the Weyl operator $W_{1,0}$ to the third (second) subsystem. Here the seed GHZ-state $GHZ_{0000}$ is recursively given by $\ket{\phi_n}=\frac{1}{\sqrt{2}} \sum^1_{i=0} \left( \mathbb{1}^{\otimes (n-1)} \otimes W_{i,i}  \right) \ket{i} \otimes \ket{\phi_{n-1}}$ with $|\phi_1\rangle=|0\rangle$ ~\cite{beatrix3} and connected to the GHZ-state in the computational basis, Eq.(\ref{GHZ}), by the unitary operation $U=(\frac{1}{2}(\mathbb{1}+i\; W_{0,1}))^{\otimes 4}$.}
\label{schemaGHZ}
\end{figure}
In this work we have focused on four-qubit GHZ-states~\cite{GHZorigin} (which is identical in this case with a graph state~\cite{graph}), having the form
\begin{equation}\label{GHZ}
\ket{GHZ_{0000}}=\frac{1}{\sqrt{2}}\left\{|0000\rangle+|1111\rangle\right\}\;.
\end{equation}
By using a construction on a minimal set of local basis rotations we have produced all $16$ orthogonal basis states.
The labelling $\ket{GHZ_{0000}}$ refers to a recursive construction method allowing to construct orthonormal basis sets of GHZ-states for any number of subsystems and any degrees of freedom introduced in Ref.~\cite{beatrix3}. A further benefit of this choice of such a construction --that we exploit in the following-- is that the geometry between local realisations of the complete basis set become transparent (see also Fig.~\ref{schemaGHZ}): let us choose without loss of generality a particular GHZ-state (such as the one given in Eq.~(\ref{GHZ})), then obviously one obtains an orthogonal state by changing the relative sign in the superposition. Wishing to construct other basis state orthogonal to these two one can apply shift operators ($|0\rangle\longrightarrow|1\rangle$ and vice versa) or/and phase operators in one or more subsystems. However, not all possibilities are successful: there exists a certain local substructure based on entangled entanglement~\cite{entangled1996,experimentangled}, which is depicted in Fig.~\ref{schemaGHZ}. Here we used the unitary Weyl operators (shift and phase operations) which correspond in our cases to the Pauli matrices ($W_{0,0}= \mathbb{1}$, $W_{0,1}=X$, $W_{1,0}=Y$, $W_{1,1}=Z$). As visualized in Fig.~\ref{schemaGHZ} there is a ``\textit{Merry Go Round}'' structure in the local realizations of GHZ-states. Each quadrant represents states where the local operations in the fourth subsystem is applied, whereas one leaves one quadrant when applying an operator to the second or third subsystem. Applying the same operation twice one moves ``backward'' on the lattice (for higher dimensions $d$, i.e. qudits, one would move ``forward'' $d-1$ times and by applying the same operator $d$-times coming back to the initial state, for details see Ref.~\cite{beatrix1}). 

Naively, mixing any two GHZ-states of the $16$ one expects that the resulting states have the same entanglement features. Indeed that is not the case, the local information makes the difference! For instance, the mixture of any two GHZ-states does not destroy the property of genuine multipartite entanglement except when they are equally mixed (as experimentally demonstrated in this work). However, for equal mixtures the entanglement properties differ considerably and can be divided in two groups:\\
\\
\textit{Type I (``\textit{twin}'' GHZ-states):} The resulting mixed state is fully separable.\\
\\
\textit{Type II (``\textit{un-twin}'' GHZ-states):} The resulting mixed state is entangled, though no longer genuine multipartite entangled, but still tri-partite entangled.\\
\\
\textit{Type I} states occur only for a single mixture, namely if one has chosen one GHZ-state in the set there exists exactly one which erases the entanglement property, a ``\textit{twin}'' GHZ-state. This is immediately clear when considering the state defined in Eq.~(\ref{GHZ}) and the one with a relative minus sign in the superposition. An equal mixture leads to zero off diagonal elements and, consequently, to a product state. In all other cases we have four non-zero off-diagonal elements for which it is not straightforward to detect their separability properties. For that we exploit the  HMGH-framework~\cite{beatrix2} providing a set of nonlinear witnesses for detecting $k$-separability. For a given matrix $\rho$ to be $k$-separable the functions $I_k(\rho)$ (defined in the appendix, Eq.(\ref{ksep})) have to be lower or equal zero, consequently a positive value detects $k$-inseparability.

For GHZ-states the criterion $I_2$ turns out to be optimal, namely the maximal value can be reached ($I_2(\ket{GHZ})=1$), whereas it is 0 for any four-qubit Dicke-state with one excitation and $\frac{1}{2}$ for any four-qubit Dicke-state with two excitations (both states are known to be genuine multipartite entangled). Differently stated $I_2$ can be turned into an optimal witness for detecting the GHZ-type entanglement of a genuinely multipartite entangled state. For our purpose the linearised version of this witness $I_2$ is sufficient due to the high symmetry of the considered states~\cite{aplica}. Note, however, for the other witnesses $I_{3,4}$ we apply the non-linearised versions. Written in Pauli's operators the linear witness detecting genuine multipartite entanglement becomes
\begin{eqnarray}\label{witness}
\tilde{I}_{2}(\rho)&=&\frac{1}{8}\biggl\langle XXXX-YYXX-YXYX-XYYX\nonumber\\
&&\hphantom{\frac{1}{8}\langle}-XXYY-XYXY-YXXY+YYYY\biggr\rangle_\rho\nonumber\\
&&-\frac{1}{8} \biggl\langle 7\mathbb{1}\mathbb{1}\mathbb{1}\mathbb{1}-ZZ\mathbb{1}\mathbb{1}-Z\mathbb{1}\mathbb{1}Z-Z\mathbb{1}Z\mathbb{1}\nonumber\\
&& \hphantom{-\frac{1}{8}\langle}-\mathbb{1}\mathbb{1}ZZ-\mathbb{1}Z\mathbb{1}Z-\mathbb{1}ZZ\mathbb{1}-ZZZZ\biggr\rangle_\rho
\end{eqnarray}
where we used the abbreviation $XXXX$ for $X\otimes X \otimes X\otimes X$ and so on. $\tilde{I}_{2}(\rho)$ detects genuine multipartite entanglement if it is greater than zero and gives the maximal value (equal to one) only for the GHZ-state in the representation Eq.(\ref{GHZ})  (by exploiting local unitary operations the criterion can be made optimal for any basis representation of the GHZ state).

In the following we describe the production of all orthogonal basis states and prove the genuine multipartite entanglement property by the above introduced criteria via different methods. Finally we discuss how the entanglement properties change when mixing of GHZ-states is considered and show that twin GHZ states behave differently.

\vspace{0.2cm}

{\bf Experimental generation of GHZ states.} GHZ states can be generated with different physical systems \cite{experimentangled,eightphotonghz,Pan2000,ghzions,cavity}.
\begin{figure*}[htb]
\begin{minipage}[c]{0.70\textwidth}
\includegraphics[width=1\textwidth]{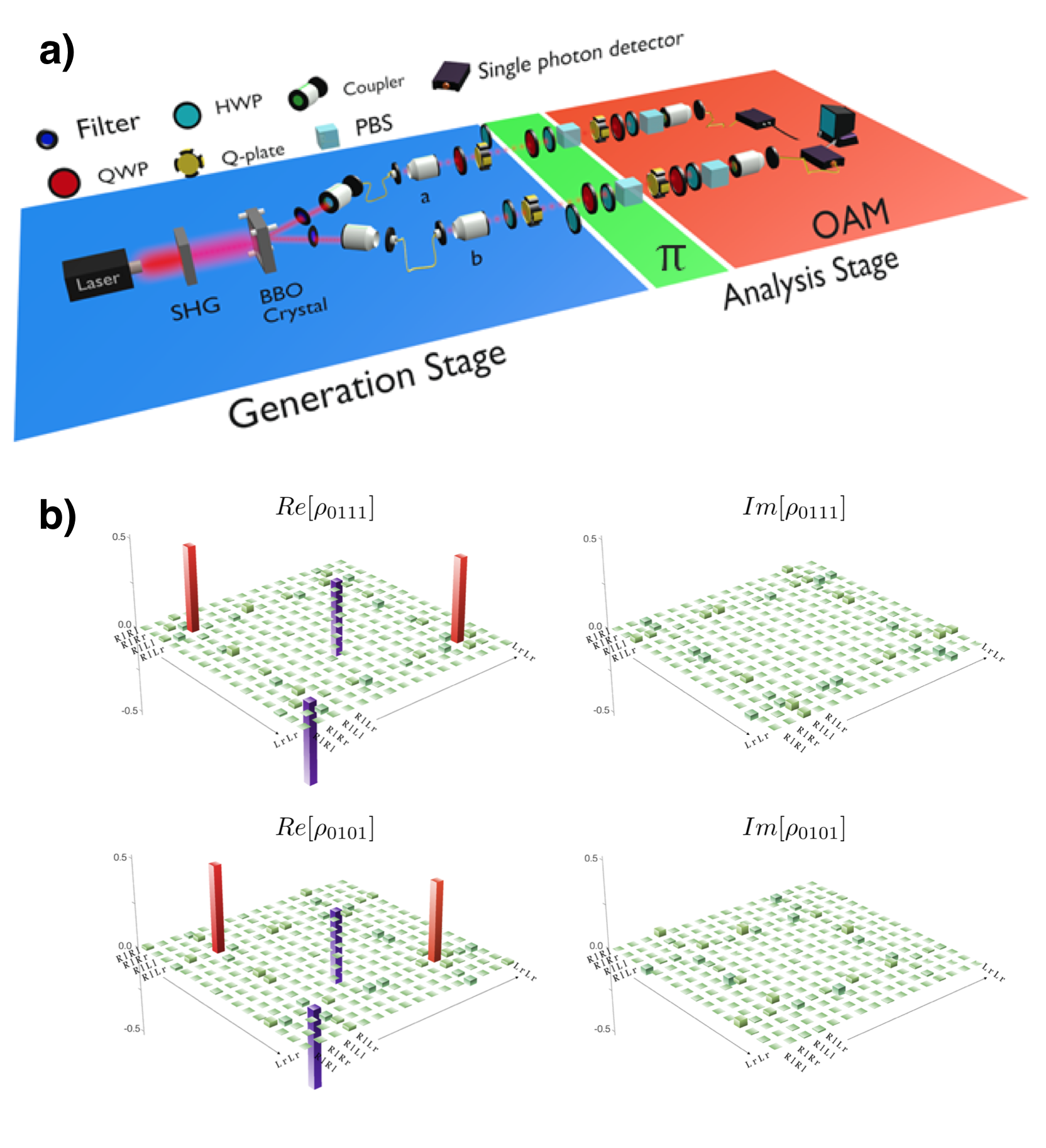}
\end{minipage}\hfill
\begin{minipage}[c]{0.30\textwidth}
\caption{Experimental setup and generated states. (a) Experimental setup for  generation and analysis of GHZ-states. In the generation stage the state of each of two entangled photons (a and b) is locally manipulated via
  QWP, HWP and q-plates with settings according to the particular GHZ state to be prepared.
The analysis stage is divided in two sections one for the polarisation analysis $\pi$  and the other for OAM analysis. The polarisation analysis is performed by using a stage composed of  QWP,  HWP and  PBS. The OAM analysis
 requires a q-plate to transfer the information encoded in the OAM space to the polarisation degree of freedom which can be then
 analysed by means of the same kit used in the $\pi$-section. After the analysis both photons are sent to single mode fibers connected to single photon detectors.
(b) Experimental density matrices of two of the generated states ($\rho_{0111}$ and $\rho_{0101}$). Real and imaginary parts of the experimental density matrices are reconstructed via full quantum state tomography.}
\label{fig:version2setup}
\end{minipage}
\end{figure*}
Here we generate photonic four-qubit GHZ states by entangling polarisation and OAM within each photon of an entangled photon pair.
To this end we exploit the q-plate \cite{Marr06,Picc10}, a birefringent slab with a suitably patterned transverse optical axis and a topological singularity at its center. Such device entangles or disentangles the OAM with the polarisation for each photon. The experimental setup is shown in Fig. \ref{fig:version2setup}(a).

The pump laser (wavelength $\lambda=397.5$ nm) is produced by a second harmonic generation (SHG) process from a Ti:saphire mode-locked laser with a repetition rate of 76-MHz. Type II spontaneous parametric down conversion (SPDC) in a $\beta$-barium borate (BBO) crystal is exploited to generate photon pairs entangled in polarisation~\cite{bbo}. These photons ($\lambda=795$ nm) are filtered in the wavelength and spatial modes by using filters with $\Delta\lambda=3$nm and single-mode fibres, respectively. The resulting state can then be written in the polarisation and OAM basis by
\begin{equation}\label{spdc}
\ket{\psi^{-}}=\frac{1}{\sqrt{2}}\{\ket{R,0}_{a}\ket{L,0}_{b}-\ket{L,0}_{a}\ket{R,0}_{b}\}\;,
\end{equation}
where $\ket{R,\ell}$ $(\ket{L,\ell})$ denotes a photon with circular right (left) polarisation and carrying $\ell\hbar$ of OAM and the subscripts $a,b$ refers to the two different photons.
Each photon is sent to a q-plate whose action is given by
\begin{equation}
\ket{R,0}+\ket{L,0}\longrightarrow \ket{L,r}+\ket{R,l}\;,
\end{equation}
where, for uniformity of notation, we wrote $r$ $(l)$ to indicate OAM eigenstates with $\ell=-1\ (+1)$. As a consequence the state (\ref{spdc}) is transformed into a GHZ-state, $\ket{GHZ_{0101}}=1/\sqrt{2}\left(\ket{RlLr}-\ket{LrRl}\right)$ (omitting the photon label subscripts). The two first qubits represent the polarisation and OAM degrees of freedom for one photon, whereas the third and fourth qubits represent the polarisation and OAM degrees of freedom for the second photon. By applying specific local transformations to $\ket{GHZ_{0101}}$ using half wave plates (HWP) and quarter wave plates (QWP) we obtain any other GHZ state of a complete set of four-qubits GHZ states.

After this stage, each photon is analyzed in the polarisation and OAM degrees of freedom. The polarisation-analysis stage is composed of QWP, HWP and polarizing beam splitter (PBS).
Since the q-plate acts as an interface between OAM and polarisation spaces, it converts the OAM-encoded information into polarisation that, in a further step, we analyze with a second polarisation analysis stage~\cite{transf1,transf2}. Finally, the photons are coupled into single mode fibers to ensure that only states with $m=0$ are detected. Our experimental setup allows thus to perform measurements of all four-qubit operators (Pauli's matrices), consequently allowing full quantum state tomography (FQST)~\cite{tomo1}.

The measurement of any four Pauli operators needs in general $16$ independent measurements. The witness given in Eq.(\ref{witness}), however, needs only $144$ measurements (not $16\cdot 16=256$ since the unit and $Z$ operator have common eigenvectors). In strong contrast, a full quantum state tomography needs $1296$ measurements.

\vspace{0.2 cm}

\begin{figure*}[htb]
\centering
\includegraphics[scale=0.4]{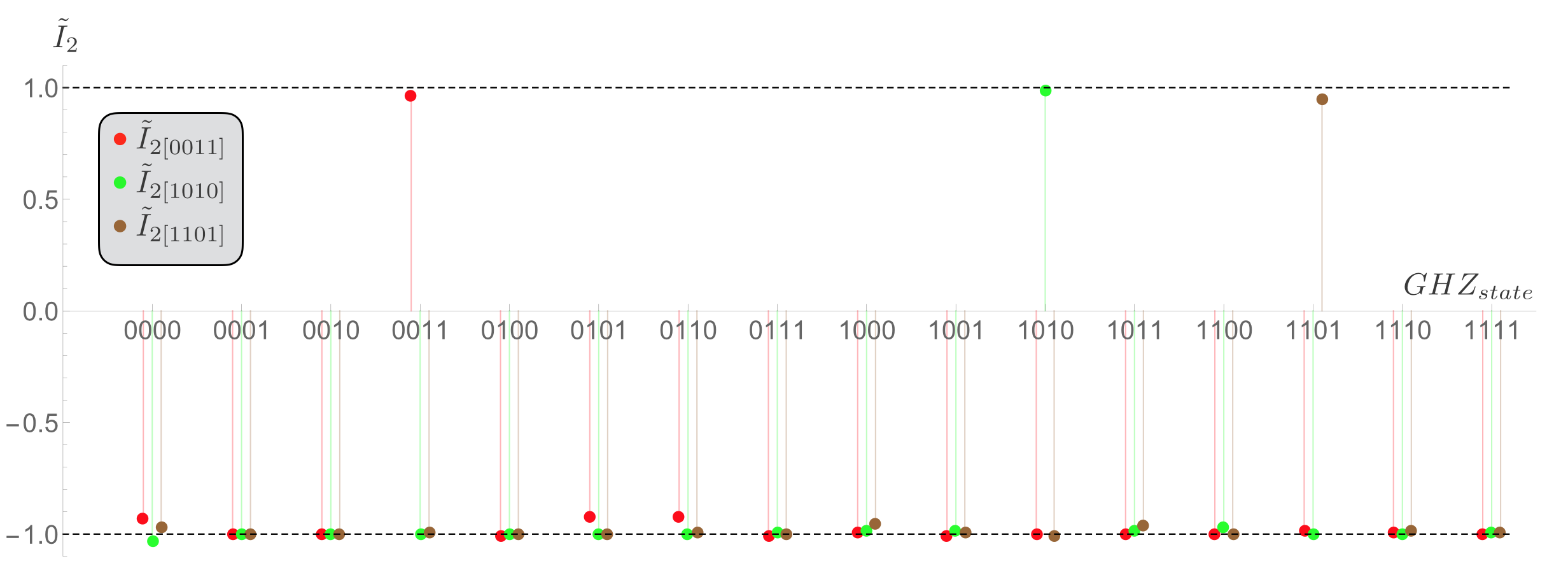}
\caption{Robustness of HMGH criterion. Application of $\tilde{I}_2$ onto all generated $GHZ$ states in the set, optimized for three different states: the twin state $\rho_{0011}$ of $\rho_{0000}$, $\rho_{1010}$ and $\rho_{1110}$. In perfect agreement with the theoretical predictions a detection is only successful in case of the matching witness. Note that for the full witness $I_2$ both twin-states are optimally detected (in linearisation the local information distinguishing the twins is lost).}
\label{3witness_tomo}
\end{figure*}

{\bf Experimental Results.}
In a first step we have generated all $16$ GHZ basis states and measured the local observables of the witness (both using raw data and with dark counts corrections). The results are listed in table~\ref{violation} and show a high stability among all $16$ GHZ states. The averaged value of the witness over all basis states is $\tilde{I}_2=0.90\pm 0.06$
 ($\tilde{I}_2=0.80\pm 0.05$)  with (without) dark counts correction, respectively.
 \begin{table}
\resizebox{8cm}{!}{
\begin{tabular}{|c||c|c|}
\hline
GHZ & $\tilde{I}_2$ & $\tilde{I}_2$ \\
State &(raw data) & (dark counts corr.)\\
\hline
$\ket{GHZ_{0000}}=\ket{RrRr}+\ket{LlLl}$&0.751$\pm$0.007&0.830$\pm$0.007\\
$\ket{GHZ_{0001}}=\ket{RrRl}-\ket{LlLr}$&0.765$\pm$0.006&0.844$\pm$0.006\\
$\ket{GHZ_{0010}}=\ket{RrRl}+\ket{LlLr}$&0.758$\pm$0.009&0.991$\pm$0.009\\
$\ket{GHZ_{0011}}=\ket{RrRr}-\ket{LlLl}$&0.871$\pm$0.003&0.966$\pm$0.003\\
$\ket{GHZ_{0100}}=\ket{RrLr}+\ket{LlRl}$&0.782$\pm$0.005&0.886 $\pm$0.005\\
$\ket{GHZ_{0101}}=\ket{RrLl}-\ket{LlRr}$&0.722$\pm$0.005&0.823$\pm$0.005\\
$\ket{GHZ_{0110}}=\ket{RrLl}+\ket{LlRr}$&0.766$\pm$0.007&0.849$\pm$0.007\\
$\ket{GHZ_{0111}}=\ket{RrLr}-\ket{LlRl}$ &0.746$\pm$0.006&0.830 $\pm$0.006\\
$\ket{GHZ_{1000}}=\ket{RlRr}+\ket{LrLl}$&0.845$\pm$0.005&0.913 $\pm$0.005\\
$\ket{GHZ_{1001}}=\ket{RlRl}-\ket{LrLr}$&0.814$\pm$0.008&0.957$\pm$0.007\\
$\ket{GHZ_{1010}}=\ket{RlRl}+\ket{LrLr}$&0.827$\pm$0.012&0.990$\pm$0.008\\
$\ket{GHZ_{1011}}=\ket{RlRr}-\ket{LrLl}$&0.763$\pm$0.006&0.838 $\pm$0.006\\
$\ket{GHZ_{1100}}=\ket{RlLr}+\ket{LrRl}$&0.827$\pm$0.004&0.915$\pm$0.004\\
$\ket{GHZ_{1101}}=\ket{RlLl} -\ket{LrRr}$&0.837$\pm$0.006&0.950 $\pm$0.006\\
$\ket{GHZ_{1110}}=\ket{RlLl}+\ket{LrRr}$&0.822$\pm$0.006&0.952 $\pm$0.006\\
$\ket{GHZ_{1111}}=\ket{RlLr}-\ket{LrRl}$&0.860$\pm$0.005&0.928$\pm$0.005\\
\hline
\end{tabular}
}
\caption{Experimental results for the witness $\tilde{I}_2$ applied to all orthogonal basis GHZ-states. Normalization factors are omitted for brevity.}
\label{violation}
\end{table} Moreover, we tested the robustness by applying three chosen witnesses to all $16$ GHZ basis states. As expected we found $\tilde{I}_2>0$ only for those states where the basis representation matches, whereas in all other cases it is clearly negative, see Fig.~\ref{3witness_tomo}.

Furthermore we have performed a full state tomography of two selected basis states (see Fig.~\ref{fig:version2setup}(b)) and applied the theoretical nonlinear witness to the obtained state $\rho_{exp}^{\text{{\tiny FQST}}}$, i.e. $I_2(\rho_{exp}^{\text{{\tiny FQST}}})$, and as well the linearised witness $\tilde{I}_2(\rho_{exp}^{\text{{\tiny FQST}}})$. The data is given in table~\ref{qrho} and shows similar results independent of the method. We checked the purity $P=Tr((\rho_{exp}^{\text{{\tiny FQST}}})^2)$ of the two states and found: $P(\rho_{0101})=0.905\pm 0.002$, $P(\rho_{0111})=0.915\pm 0.002$. The values are comparable and explain the deviations from the optimal value $\tilde{I}_{2}(GHZ)=1$.
\begin{table}[h!]
\begin{tabular}{|c|c|c||c|}
\hline
&$I_{2}(\rho_{exp}^{\text{{\tiny FQST}}})$ & $\tilde{I}_{2}(\rho_{exp}^{\text{{\tiny FQST}}})$& $\tilde{I}_{2}$ \\
\hline
$\rho_{0111}$&$0.896 \pm 0.002 $&$0.865 \pm 0.002 $& $0.830 \pm 0.006$  \\
\hline
$\rho_{0101}$&$0.893 \pm 0.002 $&$0.845 \pm 0.003$ & $0.823 \pm 0.005 $ \\
\hline
\end{tabular}
\caption{Evaluation of the HMGH criterion $I_{2}$ for two generated GHZ states. Starting from the experimental density matrix $\rho_{exp}^{\text{{\tiny FQST}}}$ we evaluated the criterion $I_2(\rho_{exp}^{\text{{\tiny FQST}}})$ directly (first column) and its linearised version, Eq.~(\ref{witness}), (second column). These two values can be compared to the values directly obtained by measuring the witness, Eq.~(\ref{witness}), (third column).}\label{qrho}
\end{table}

In summary, all produced states are certainly genuine multipartite entangled, i.e. there exist no bipartition via any partition of all involved degrees of freedom. Since all measured values are in good experimental agreement the data clearly shows the independence on the degrees of freedom chosen.

\vspace{0.7 cm}

\textbf{Entanglement properties of mixtures of GHZ states.}
For revealing the local substructure of mixtures of GHZ states we generated and measured all the sixteen pure GHZ and we summed the raw data by weighting each component proportionally to its statistical weight in the mixture.
This procedure is analogous to performing the statistical mixture realized in time, i.e. measuring each state for a time proportional to its weight in the mixture.
In particular we consider a mixed state composed of white noise and (in general) three GHZ states $\rho_i$
\begin{eqnarray}
\rho(\alpha,\beta,\gamma)=\frac{1-\alpha-\beta-\gamma}{16}\mathbb{1}+\alpha \rho_1+\beta \rho_2+\gamma \rho_3
\end{eqnarray}
where $\alpha$, $\beta$ and $\gamma$ are statistical weights. As stated in the beginning a chosen GHZ-state has always exactly one closely related geometrical twin. Without loss of generality we assume that $\rho_1$, $\rho_2$ are such a pair, i.e. the equal mixture of both states results in a separable state. Whereas a mixture of $\rho_1$ with any other GHZ-state $\rho_3$ is not $k=3$-separable.

\begin{figure*}
\centering
\includegraphics[width=1\textwidth]{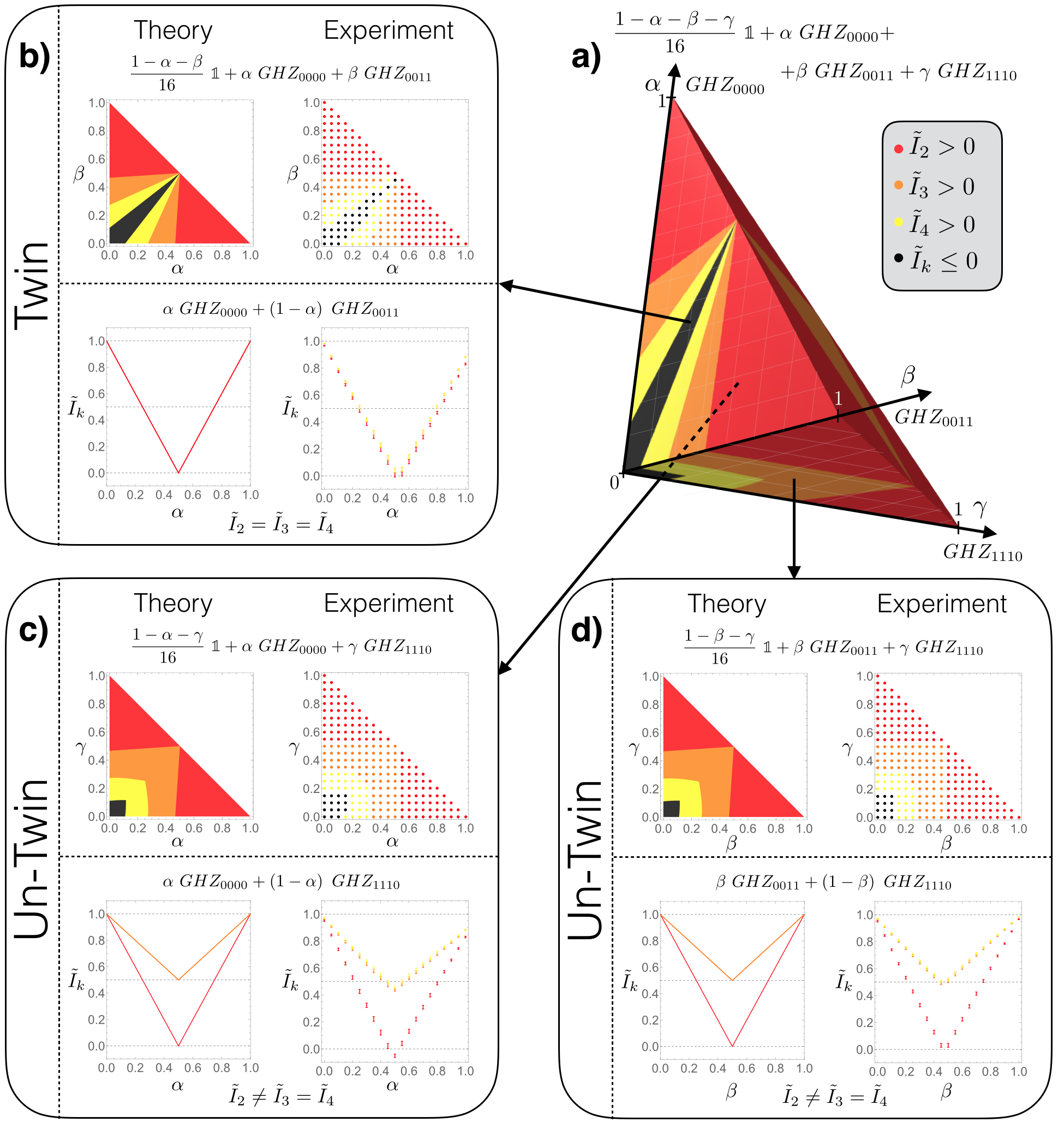}
\caption{Theoretical and experimental results for GHZ mixtures. (a) Theoretical geometry of the mixture of three GHZ in the presence of white noise. The parameters $\alpha$ and $\beta$ are the statistical weights of two twin GHZ (in this case $GHZ_{0000}$ and $GHZ_{0011}$) while $\gamma$ is the weight of the un-twin one ($GHZ_{1110}$).
Red regions represent mixed states which are not bi-separable (i.e. are entangled in a multipartite sense), orange (yellow) regions correspond to states which are not $k=3$ ($k=4$)-separable but are bi-separable, black regions represent those states on which $I_k \leq 0$ or equivalently the states are invariant under partial transpose.
This peculiar geometry holds for any choice of two twin GHZ and an un-twin one.
(b-d) Theoretical and experimental geometry for the mixtures of two GHZ with and without white noise.
On the left side of each box are shown the theoretical mixtures with (on the top) and without (on the bottom) noise, while on the right are shown the corresponding experimental results.
The (b) box shows mixtures of twin GHZ, where the equal mixture of both states results in a separable state.
(c,d) boxes show mixtures of two pairs of un-twin GHZ having the same geometry: equal mixture of both states are bi-separable but are not $k=3$-separable.}
\label{geometry}
\end{figure*}

Fig.~\ref{geometry} shows the theoretical and experimental geometry for a given choice of $\rho(\alpha,\beta,\gamma)$ (section (a)) and its corresponding sub-mixtures of two GHZ with (and without) white noise (sections (b-d)): $\rho(\alpha,\beta)$, $\rho(\alpha,\gamma)$ and $\rho(\beta,\gamma)$.
This figure shows clearly how mixtures of twin and un-twin GHZ exhibit different behaviours: mixtures of twin pairs (b) are fully separable if weights in the mixture have the same value, while this is not true if we look at mixtures of un-twin pairs (c,d) in which the states are bi-separable but not three-separable considering again mixtures having the same weights for both the states.
Finally one can notice that the regions of bi-inseparability coincide for twin or un-twin mixtures, although regions of three and four-inseparability are different in the two cases. Moreover, looking separately to twin and un-twin mixtures, three and four-inseparability coincide in absence of noise, while they show a different behaviour when the mixture becomes noisy.

\vspace{0.2cm}
\textbf{Discussions and Outlook}

We have considered states in a four-tensored Hilbert-space where each subspace is described by two dimensions which we physically achieved by manipulating the polarisation and orbital momentum degrees of freedom of two photons. Producing a complete set of orthogonal GHZ-states and their detection via entanglement witnesses showed a high quality in always achieving states with same entanglement properties but locally different geometries. This local differences are important when mixing those states. In particular we proved experimentally that among the $16$ GHZ-states each GHZ-state has always a twin that when mixed with equal weights gives a fully separable state. In opposition to any other balanced mixtures of GHZ-states destroying genuine multipartite entanglement, but not any other type of entanglement.

This property is important for example in secret sharing protocols based on the mixtures of GHZ-states~\cite{aplica} and for quantum algorithms exploring different types of multipartite entanglement~\cite{Jozsa2011,macchiavello}. Certainly, this local information beween orthogonal basis states is relevant for any experimental setup and can be exploit to generate particular types of entanglement.

\vspace{0.6cm}
{\bf Acknowledgments:} We thank L. Marrucci and B. Piccirillo for providing q-plates and for useful discussions. B.C.H. acknowledges support from the Austrian Science Fund (FWF 23627). G.C. thanks Becas Chile and Conicyt for a doctoral fellowship.
This work was supported by PRIN (Programmi di ricerca di rilevante interesse nazionale) project AQUASIM and ERC-Starting Grant 3D-QUEST (3D-Quantum Integrated Optical Simulation; Grant Agreement No. 307783): www.3dquest.eu.

\section{Methods}

\textbf{The $k$-separability criteria}: 
In Ref.~\cite{beatrix4} it was proven that any state $\rho$ (mixed or pure) that is $k$-separable has to satisfy $I_k(\rho)\leq 0$ where for a $16\times 16$ matrix  $\rho$ the functions (optimized for the state in eq. \eqref{GHZ}) read explicitly
\begin{eqnarray}\label{ksep}
I_{k=2}(\rho)&=&2 |\rho_{1,16}|-\left(\sqrt{\rho_{2,2}\rho_{15,15}}+\sqrt{\rho_{3,3}\rho_{14,14}}\right.\nonumber\\
&&+\sqrt{\rho_{4,4}\rho_{13,13}}+\sqrt{\rho_{5,5}\rho_{12,12}}+\sqrt{\rho_{6,6}\rho_{11,11}}\nonumber\\
&&\left.+\sqrt{\rho_{7,7}\rho_{10,10}}+\sqrt{\rho_{8,8}\rho_{9,9}}\right)\;,\nonumber\\
I_{k=3}(\rho)&=&2 |\rho_{1,16}|-\left((\rho_{2,2}\rho_{3,3}\rho_{4,4}\rho_{13,13}\rho_{14,14}\rho_{15,15})^{\frac{1}{6}}\right.\nonumber\\
&&+(\rho_{2,2}\rho_{5,5}\rho_{6,6}\rho_{11,11}\rho_{12,12}\rho_{15,15})^{\frac{1}{6}}\nonumber\\
&&+(\rho_{2,2}\rho_{7,7}\rho_{8,8}\rho_{9,9}\rho_{10,10}\rho_{15,15})^{\frac{1}{6}}\nonumber\\
&&+(\rho_{3,3}\rho_{5,5}\rho_{7,7}\rho_{10,10}\rho_{12,12}\rho_{14,14})^{\frac{1}{6}}\nonumber\\
&&+(\rho_{3,3}\rho_{6,6}\rho_{8,8}\rho_{9,9}\rho_{11,11}\rho_{14,14})^{\frac{1}{6}}\nonumber\\
&&\left.+(\rho_{4,4}\rho_{5,5}\rho_{8,8}\rho_{9,9}\rho_{12,12}\rho_{13,13})^{\frac{1}{6}}\right)\;,\nonumber\\
I_{k=4}(\rho)&=&2 |\rho_{1,16}|-2 (\rho_{22} \rho_{33} \rho_{55} \rho_{88} \rho_{99} \rho_{12,12} \rho_{14,14} \rho_{15,15})^{\frac{1}{8}}\;.\nonumber\\
\end{eqnarray}


\newpage 

\begin{thebibliography}{10}
\expandafter\ifx\csname url\endcsname\relax
  \def\url#1{\texttt{#1}}\fi
\expandafter\ifx\csname urlprefix\endcsname\relax\def\urlprefix{}\fi
\providecommand{\bibinfo}[2]{#2}
\providecommand{\eprint}[2][]{\url{#2}}

\bibitem{Reventanglem}
\bibinfo{author}{Horodecki, R.}, \bibinfo{author}{Horodecki, P.},
  \bibinfo{author}{Horodecki, M.} \& \bibinfo{author}{Horodecki, K.}
\newblock \bibinfo{title}{Quantum entanglement}.
\newblock \emph{\bibinfo{journal}{Rev. Mod. Phys.}}
  \textbf{\bibinfo{volume}{81}}, \bibinfo{pages}{865--942}
  (\bibinfo{year}{2009}).
\newblock \urlprefix\url{http://link.aps.org/doi/10.1103/RevModPhys.81.865}.

\bibitem{pub112}
\bibinfo{author}{Zukowski, M.} \& \bibinfo{author}{Zeilinger, A.}
\newblock \bibinfo{title}{Test of the {B}ell's inequality based on phase and
  linear momentum as well as spin}.
\newblock \emph{\bibinfo{journal}{Phys. Lett. A}}
  \textbf{\bibinfo{volume}{155}}, \bibinfo{pages}{69} (\bibinfo{year}{1991}).
\newblock
  \urlprefix\url{http://www.sciencedirect.com/science/article/pii/037596019190566Q}.

\bibitem{carlos}
\bibinfo{author}{Neves, L.}, \bibinfo{author}{Lima, G.},
  \bibinfo{author}{Delgado, A.} \& \bibinfo{author}{Saavedra, C.}
\newblock \bibinfo{title}{Hybrid photonic entanglement: Realization,
  characterization, and applications}.
\newblock \emph{\bibinfo{journal}{Phys. Rev. A}} \textbf{\bibinfo{volume}{80}},
  \bibinfo{pages}{042322} (\bibinfo{year}{2009}).
\newblock \urlprefix\url{http://link.aps.org/doi/10.1103/PhysRevA.80.042322}.

\bibitem{Pan}
\bibinfo{author}{Pan, J.~W.} \emph{et~al.}
\newblock \bibinfo{title}{Multiphoton entanglement and interferometry}.
\newblock \emph{\bibinfo{journal}{Rev. Mod. Phys.}}
  \textbf{\bibinfo{volume}{84}}, \bibinfo{pages}{777--838}
  (\bibinfo{year}{2012}).
\newblock \urlprefix\url{http://link.aps.org/doi/10.1103/RevModPhys.84.777}.

\bibitem{belldof}
\bibinfo{author}{Ma, X.~S.}, \bibinfo{author}{Qarry, A.},
  \bibinfo{author}{Kofler, J.}, \bibinfo{author}{Jennewein, T.} \&
  \bibinfo{author}{Zeilinger, A.}
\newblock \bibinfo{title}{Experimental violation of a {B}ell inequality with
  two different degrees of freedom of entangled particle pairs}.
\newblock \emph{\bibinfo{journal}{Phys. Rev. A}} \textbf{\bibinfo{volume}{79}},
  \bibinfo{pages}{042101} (\bibinfo{year}{2009}).
\newblock \urlprefix\url{http://link.aps.org/doi/10.1103/PhysRevA.79.042101}.

\bibitem{nagasanta}
\bibinfo{author}{Nagali, E.} \emph{et~al.}
\newblock \bibinfo{title}{Optimal quantum cloning of orbital angular momentum
  photon qubits through hong-ou-mandel coalescence}.
\newblock \emph{\bibinfo{journal}{Nat Photon}} \textbf{\bibinfo{volume}{3}},
  \bibinfo{pages}{720--723} (\bibinfo{year}{2009}).
\newblock \urlprefix\url{http://dx.doi.org/10.1038/nphoton.2009.214}.

\bibitem{kwiatbarr}
\bibinfo{author}{Barreiro, J.~T.}, \bibinfo{author}{Wei, T.-C.} \&
  \bibinfo{author}{Kwiat, P.~G.}
\newblock \bibinfo{title}{Remote preparation of single-photon ``hybrid''
  entangled and vector-polarization states}.
\newblock \emph{\bibinfo{journal}{Phys. Rev. Lett.}}
  \textbf{\bibinfo{volume}{105}}, \bibinfo{pages}{030407}
  (\bibinfo{year}{2010}).
\newblock
  \urlprefix\url{http://link.aps.org/doi/10.1103/PhysRevLett.105.030407}.

\bibitem{aielloleuchs}
\bibinfo{author}{Aiello, A.}, \bibinfo{author}{T{\"o}ppel, F.},
  \bibinfo{author}{Marquardt, C.}, \bibinfo{author}{Giacobino, E.} \&
  \bibinfo{author}{Leuchs, G.}
\newblock \bibinfo{title}{Quantum-like nonseparable structures in optical
  beams}.
\newblock \emph{\bibinfo{journal}{New Journal of Physics}}
  \textbf{\bibinfo{volume}{17}}, \bibinfo{pages}{043024}
  (\bibinfo{year}{2015}).
\newblock \urlprefix\url{http://stacks.iop.org/1367-2630/17/i=4/a=043024}.

\bibitem{detchall}
\bibinfo{author}{Hiesmayr, B.~C.}, \bibinfo{author}{Huber, M.} \&
  \bibinfo{author}{Krammer, P.}
\newblock \bibinfo{title}{Two computable sets of multipartite entanglement
  measures}.
\newblock \emph{\bibinfo{journal}{Phys. Rev. A}} \textbf{\bibinfo{volume}{79}},
  \bibinfo{pages}{062308} (\bibinfo{year}{2009}).
\newblock \urlprefix\url{http://link.aps.org/doi/10.1103/PhysRevA.79.062308}.

\bibitem{Guhne20091}
\bibinfo{author}{G{\"u}hne, O.} \& \bibinfo{author}{T{\'o}th, G.}
\newblock \bibinfo{title}{Entanglement detection}.
\newblock \emph{\bibinfo{journal}{Physics Reports}}
  \textbf{\bibinfo{volume}{474}}, \bibinfo{pages}{1 -- 75}
  (\bibinfo{year}{2009}).
\newblock
  \urlprefix\url{http://www.sciencedirect.com/science/article/pii/S0370157309000623}.

\bibitem{Bennet}
\bibinfo{author}{Bennett, C.~H.} \emph{et~al.}
\newblock \bibinfo{title}{Teleporting an unknown quantum state via dual
  classical and {E}instein-{P}odolsky-{R}osen channels}.
\newblock \emph{\bibinfo{journal}{Phys. Rev. Lett.}}
  \textbf{\bibinfo{volume}{70}}, \bibinfo{pages}{1895--1899}
  (\bibinfo{year}{1993}).
\newblock \urlprefix\url{http://link.aps.org/doi/10.1103/PhysRevLett.70.1895}.

\bibitem{dense}
\bibinfo{author}{Bennett, C.~H.} \& \bibinfo{author}{Wiesner, S.~J.}
\newblock \bibinfo{title}{Communication via one- and two-particle operators on
  einstein-podolsky-rosen states}.
\newblock \emph{\bibinfo{journal}{Phys. Rev. Lett.}}
  \textbf{\bibinfo{volume}{69}}, \bibinfo{pages}{2881--2884}
  (\bibinfo{year}{1992}).
\newblock \urlprefix\url{http://link.aps.org/doi/10.1103/PhysRevLett.69.2881}.

\bibitem{shor1995}
\bibinfo{author}{Shor, P.~W.}
\newblock \bibinfo{title}{Scheme for reducing decoherence in quantum computer
  memory}.
\newblock \emph{\bibinfo{journal}{Phys. Rev. A}} \textbf{\bibinfo{volume}{52}},
  \bibinfo{pages}{R2493--R2496} (\bibinfo{year}{1995}).
\newblock \urlprefix\url{http://link.aps.org/doi/10.1103/PhysRevA.52.R2493}.

\bibitem{steane1996}
\bibinfo{author}{Steane, A.~M.}
\newblock \bibinfo{title}{Error correcting codes in quantum theory}.
\newblock \emph{\bibinfo{journal}{Phys. Rev. Lett.}}
  \textbf{\bibinfo{volume}{77}}, \bibinfo{pages}{793--797}
  (\bibinfo{year}{1996}).
\newblock \urlprefix\url{http://link.aps.org/doi/10.1103/PhysRevLett.77.793}.

\bibitem{feynman1982}
\bibinfo{author}{Feynman, R.~P.}
\newblock \bibinfo{title}{Simulating physics with computers}.
\newblock \emph{\bibinfo{journal}{International journal of theoretical
  physics}} \textbf{\bibinfo{volume}{21}}, \bibinfo{pages}{467--488}
  (\bibinfo{year}{1982}).

\bibitem{ekert1991}
\bibinfo{author}{Ekert, A.~K.}
\newblock \bibinfo{title}{Quantum cryptography based on {B}ell's theorem}.
\newblock \emph{\bibinfo{journal}{Phys. Rev. Lett.}}
  \textbf{\bibinfo{volume}{67}}, \bibinfo{pages}{661--663}
  (\bibinfo{year}{1991}).
\newblock \urlprefix\url{http://link.aps.org/doi/10.1103/PhysRevLett.67.661}.

\bibitem{QKDE}
\bibinfo{author}{Jennewein, T.}, \bibinfo{author}{Simon, C.},
  \bibinfo{author}{Weihs, G.}, \bibinfo{author}{Weinfurter, H.} \&
  \bibinfo{author}{Zeilinger, A.}
\newblock \bibinfo{title}{Quantum cryptography with entangled photons}.
\newblock \emph{\bibinfo{journal}{Phys. Rev. Lett.}}
  \textbf{\bibinfo{volume}{84}}, \bibinfo{pages}{4729--4732}
  (\bibinfo{year}{2000}).
\newblock \urlprefix\url{http://link.aps.org/doi/10.1103/PhysRevLett.84.4729}.

\bibitem{Car12}
\bibinfo{author}{Cardano, F.} \emph{et~al.}
\newblock \bibinfo{title}{Polarization pattern of vector vortex beams generated
  by q-plates with different topological charges}.
\newblock \emph{\bibinfo{journal}{Appl. Opt.}} \textbf{\bibinfo{volume}{51}},
  \bibinfo{pages}{C1--C6} (\bibinfo{year}{2012}).
\newblock \urlprefix\url{http://ao.osa.org/abstract.cfm?URI=ao-51-10-C1}.

\bibitem{sing}
\bibinfo{author}{Dennis, M.~R.}, \bibinfo{author}{O'Holleran, K.} \&
  \bibinfo{author}{Padgett, M.~J.}
\newblock \bibinfo{title}{Singular optics: Optical vortices and polarization
  singularities}.
\newblock \emph{\bibinfo{journal}{Prog. Opt.}} \textbf{\bibinfo{volume}{53}},
  \bibinfo{pages}{293--363} (\bibinfo{year}{2009}).

\bibitem{vazirioam}
\bibinfo{author}{Mair, A.}, \bibinfo{author}{Vaziri, A.},
  \bibinfo{author}{Weihs, G.} \& \bibinfo{author}{Zeilinger, A.}
\newblock \bibinfo{title}{Entanglement of the orbital angular momentum states
  of photons}.
\newblock \emph{\bibinfo{journal}{Nature}} \textbf{\bibinfo{volume}{412}},
  \bibinfo{pages}{313--316} (\bibinfo{year}{2001}).
\newblock \urlprefix\url{http://dx.doi.org/10.1038/35085529}.

\bibitem{Nagali10}
\bibinfo{author}{Nagali, E.} \& \bibinfo{author}{Sciarrino, F.}
\newblock \bibinfo{title}{Generation of hybrid polarization-orbital angular
  momentum entangled states}.
\newblock \emph{\bibinfo{journal}{Opt. Express}} \textbf{\bibinfo{volume}{18}},
  \bibinfo{pages}{18243--18248} (\bibinfo{year}{2010}).
\newblock
  \urlprefix\url{http://www.opticsexpress.org/abstract.cfm?URI=oe-18-17-18243}.

\bibitem{marrpadg}
\bibinfo{author}{Karimi, E.} \emph{et~al.}
\newblock \bibinfo{title}{Spin-orbit hybrid entanglement of photons and quantum
  contextuality}.
\newblock \emph{\bibinfo{journal}{Phys. Rev. A}} \textbf{\bibinfo{volume}{82}},
  \bibinfo{pages}{022115} (\bibinfo{year}{2010}).
\newblock \urlprefix\url{http://link.aps.org/doi/10.1103/PhysRevA.82.022115}.

\bibitem{Torres}
\bibinfo{author}{Vall\'es, A.} \emph{et~al.}
\newblock \bibinfo{title}{Generation of tunable entanglement and violation of a
  {B}ell-like inequality between different degrees of freedom of a single
  photon}.
\newblock \emph{\bibinfo{journal}{Phys. Rev. A}} \textbf{\bibinfo{volume}{90}},
  \bibinfo{pages}{052326} (\bibinfo{year}{2014}).
\newblock \urlprefix\url{http://link.aps.org/doi/10.1103/PhysRevA.90.052326}.

\bibitem{cabello}
\bibinfo{author}{D'Ambrosio, V.} \emph{et~al.}
\newblock \bibinfo{title}{Experimental implementation of a {K}ochen-{S}pecker
  set of quantum tests}.
\newblock \emph{\bibinfo{journal}{Phys. Rev. X}} \textbf{\bibinfo{volume}{3}},
  \bibinfo{pages}{011012} (\bibinfo{year}{2013}).
\newblock \urlprefix\url{http://link.aps.org/doi/10.1103/PhysRevX.3.011012}.

\bibitem{VVB}
\bibinfo{author}{D'Ambrosio, V.} \emph{et~al.}
\newblock \bibinfo{title}{Hybrid entangled entanglement in vector vortex
  beams}.
\newblock \emph{\bibinfo{journal}{arXiv [quant-ph]}}  (\bibinfo{year}{2015}).
\newblock \urlprefix\url{http://arxiv.org/abs/1507.08887v1}.

\bibitem{witnessblatt}
\bibinfo{author}{Barreiro, J.~T.} \emph{et~al.}
\newblock \bibinfo{title}{Demonstration of genuine multipartite entanglement
  with device-independent witnesses}.
\newblock \emph{\bibinfo{journal}{Nat Phys}} \textbf{\bibinfo{volume}{9}},
  \bibinfo{pages}{559--562} (\bibinfo{year}{2013}).
\newblock \urlprefix\url{http://dx.doi.org/10.1038/nphys2705}.

\bibitem{beatrix2}
\bibinfo{author}{Huber, M.}, \bibinfo{author}{Mintert, F.},
  \bibinfo{author}{Gabriel, A.} \& \bibinfo{author}{Hiesmayr, B.~C.}
\newblock \bibinfo{title}{Detection of high-dimensional genuine multipartite
  entanglement of mixed states}.
\newblock \emph{\bibinfo{journal}{Phys. Rev. Lett.}}
  \textbf{\bibinfo{volume}{104}}, \bibinfo{pages}{210501}
  (\bibinfo{year}{2010}).
\newblock
  \urlprefix\url{http://link.aps.org/doi/10.1103/PhysRevLett.104.210501}.

\bibitem{beatrix3}
\bibinfo{author}{Uchida, G.}, \bibinfo{author}{Bertlmann, R.~A.} \&
  \bibinfo{author}{Hiesmayr, B.~C.}
\newblock \bibinfo{title}{Entangled entanglement: A construction procedure}.
\newblock \emph{\bibinfo{journal}{Physics Letters A}}
  \textbf{\bibinfo{volume}{379}}, \bibinfo{pages}{2698 -- 2703}
  (\bibinfo{year}{2015}).
\newblock
  \urlprefix\url{http://www.sciencedirect.com/science/article/pii/S0375960115006775}.

\bibitem{GHZorigin}
\bibinfo{author}{Greenberger, D.~M.}, \bibinfo{author}{Horne, M.~A.} \&
  \bibinfo{author}{Zeilinger, A.}
\newblock \bibinfo{title}{Going beyond bell'Äôs theorem}.
\newblock \emph{\bibinfo{journal}{Bell'Äôs Theorem, Quantum Theory, and
  Conceptions of the Universe}} \bibinfo{pages}{69--72} (\bibinfo{year}{1989}).

\bibitem{graph}
\bibinfo{author}{Hein, M.}, \bibinfo{author}{Eisert, J.} \&
  \bibinfo{author}{Briegel, H.~J.}
\newblock \bibinfo{title}{Multiparty entanglement in graph states}.
\newblock \emph{\bibinfo{journal}{Phys. Rev. A}} \textbf{\bibinfo{volume}{69}},
  \bibinfo{pages}{062311} (\bibinfo{year}{2004}).
\newblock \urlprefix\url{http://link.aps.org/doi/10.1103/PhysRevA.69.062311}.

\bibitem{entangled1996}
\bibinfo{author}{Krenn, G.} \& \bibinfo{author}{Zeilinger, A.}
\newblock \bibinfo{title}{Entangled entanglement}.
\newblock \emph{\bibinfo{journal}{Phys. Rev. A}} \textbf{\bibinfo{volume}{54}},
  \bibinfo{pages}{1793--1797} (\bibinfo{year}{1996}).
\newblock \urlprefix\url{http://link.aps.org/doi/10.1103/PhysRevA.54.1793}.

\bibitem{experimentangled}
\bibinfo{author}{Walther, P.}, \bibinfo{author}{Resch, K.~J.},
  \bibinfo{author}{Brukner, C.} \& \bibinfo{author}{Zeilinger, A.}
\newblock \bibinfo{title}{Experimental entangled entanglement}.
\newblock \emph{\bibinfo{journal}{Phys. Rev. Lett.}}
  \textbf{\bibinfo{volume}{97}}, \bibinfo{pages}{020501}
  (\bibinfo{year}{2006}).
\newblock
  \urlprefix\url{http://link.aps.org/doi/10.1103/PhysRevLett.97.020501}.

\bibitem{beatrix1}
\bibinfo{author}{Uchida, G.}, \bibinfo{author}{Reinhold, G.},
  \bibinfo{author}{Bertlmann, A.} \& \bibinfo{author}{Hiesmayr, B.~C.}
\newblock \bibinfo{title}{Entangled entanglement: The geometry of {GHZ}
  states}.
\newblock \emph{\bibinfo{journal}{arXiv [quant-ph]}}  (\bibinfo{year}{2014}).
\newblock \urlprefix\url{http://arxiv.org/abs/1410.7145v1}.

\bibitem{aplica}
\bibinfo{author}{Schauer, S.}, \bibinfo{author}{Huber, M.} \&
  \bibinfo{author}{Hiesmayr, B.~C.}
\newblock \bibinfo{title}{Experimentally feasible security check for $n$-qubit
  quantum secret sharing}.
\newblock \emph{\bibinfo{journal}{Phys. Rev. A}} \textbf{\bibinfo{volume}{82}},
  \bibinfo{pages}{062311} (\bibinfo{year}{2010}).
\newblock \urlprefix\url{http://link.aps.org/doi/10.1103/PhysRevA.82.062311}.

\bibitem{eightphotonghz}
\bibinfo{author}{Huang, Y.-F.} \emph{et~al.}
\newblock \bibinfo{title}{Experimental generation of an eight-photon
  {G}reenberger-{H}orne-{Z}eilinger state}.
\newblock \emph{\bibinfo{journal}{Nat Commun}} \textbf{\bibinfo{volume}{2}},
  \bibinfo{pages}{546} (\bibinfo{year}{2011}).
\newblock \urlprefix\url{http://dx.doi.org/10.1038/ncomms1556}.

\bibitem{Pan2000}
\bibinfo{author}{Pan, J.~W.}, \bibinfo{author}{Bouwmeester, D.},
  \bibinfo{author}{Daniell, M.}, \bibinfo{author}{Weinfurter, H.} \&
  \bibinfo{author}{Zeilinger, A.}
\newblock \bibinfo{title}{Experimental test of quantum nonlocality in
  three-photon {G}reenberger-{H}orne-{Z}eilinger entanglement}.
\newblock \emph{\bibinfo{journal}{Nature}} \textbf{\bibinfo{volume}{403}},
  \bibinfo{pages}{515--519} (\bibinfo{year}{2000}).
\newblock \urlprefix\url{http://dx.doi.org/10.1038/35000514}.

\bibitem{ghzions}
\bibinfo{author}{Leibfried, D.} \emph{et~al.}
\newblock \bibinfo{title}{Toward heisenberg-limited spectroscopy with
  multiparticle entangled states}.
\newblock \emph{\bibinfo{journal}{Science}} \textbf{\bibinfo{volume}{304}},
  \bibinfo{pages}{1476--1478} (\bibinfo{year}{2004}).
\newblock
  \urlprefix\url{http://www.sciencemag.org/content/304/5676/1476.abstract}.

\bibitem{cavity}
\bibinfo{author}{Brattke, S.}, \bibinfo{author}{Varcoe, B. T.~H.} \&
  \bibinfo{author}{Walther, H.}
\newblock \bibinfo{title}{Generation of photon number states on demand via
  cavity quantum electrodynamics}.
\newblock \emph{\bibinfo{journal}{Phys. Rev. Lett.}}
  \textbf{\bibinfo{volume}{86}}, \bibinfo{pages}{3534--3537}
  (\bibinfo{year}{2001}).
\newblock \urlprefix\url{http://link.aps.org/doi/10.1103/PhysRevLett.86.3534}.

\bibitem{Marr06}
\bibinfo{author}{Marrucci, L.}, \bibinfo{author}{Manzo, C.} \&
  \bibinfo{author}{Paparo, D.}
\newblock \bibinfo{title}{Optical spin-to-orbital angular momentum conversion
  in inhomogeneous anisotropic media}.
\newblock \emph{\bibinfo{journal}{Phys. Rev. Lett.}}
  \textbf{\bibinfo{volume}{96}}, \bibinfo{pages}{163905}
  (\bibinfo{year}{2006}).
\newblock
  \urlprefix\url{http://link.aps.org/doi/10.1103/PhysRevLett.96.163905}.

\bibitem{Picc10}
\bibinfo{author}{Piccirillo, B.}, \bibinfo{author}{D'~Ambrosio, V.},
  \bibinfo{author}{Slussarenko, S.}, \bibinfo{author}{Marrucci, L.} \&
  \bibinfo{author}{Santamato, E.}
\newblock \bibinfo{title}{Photon spin-to-orbital angular momentum conversion
  via an electrically tunable q-plate}.
\newblock \emph{\bibinfo{journal}{Applied Physics Letters}}
  \textbf{\bibinfo{volume}{97}}, \bibinfo{pages}{--} (\bibinfo{year}{2010}).
\newblock
  \urlprefix\url{http://scitation.aip.org/content/aip/journal/apl/97/24/10.1063/1.3527083}.

\bibitem{bbo}
\bibinfo{author}{Kwiat, P.~G.} \emph{et~al.}
\newblock \bibinfo{title}{New high-intensity source of polarization-entangled
  photon pairs}.
\newblock \emph{\bibinfo{journal}{Phys. Rev. Lett.}}
  \textbf{\bibinfo{volume}{75}}, \bibinfo{pages}{4337--4341}
  (\bibinfo{year}{1995}).
\newblock \urlprefix\url{http://link.aps.org/doi/10.1103/PhysRevLett.75.4337}.

\bibitem{transf1}
\bibinfo{author}{Nagali, E.} \emph{et~al.}
\newblock \bibinfo{title}{Quantum information transfer from spin to orbital
  angular momentum of photons}.
\newblock \emph{\bibinfo{journal}{Phys. Rev. Lett.}}
  \textbf{\bibinfo{volume}{103}}, \bibinfo{pages}{013601}
  (\bibinfo{year}{2009}).
\newblock
  \urlprefix\url{http://link.aps.org/doi/10.1103/PhysRevLett.103.013601}.

\bibitem{transf2}
\bibinfo{author}{D'Ambrosio, V.} \emph{et~al.}
\newblock \bibinfo{title}{Deterministic qubit transfer between orbital and spin
  angular momentum of single photons}.
\newblock \emph{\bibinfo{journal}{Opt. Lett.}} \textbf{\bibinfo{volume}{37}},
  \bibinfo{pages}{172--174} (\bibinfo{year}{2012}).
\newblock \urlprefix\url{http://ol.osa.org/abstract.cfm?URI=ol-37-2-172}.

\bibitem{tomo1}
\bibinfo{author}{Altepeter, J.}, \bibinfo{author}{Jeffrey, E.} \&
  \bibinfo{author}{Kwiat, P.}
\newblock \bibinfo{title}{Photonic state tomography}.
\newblock vol.~\bibinfo{volume}{52} of \emph{\bibinfo{series}{Advances In
  Atomic, Molecular, and Optical Physics}}, \bibinfo{pages}{105 -- 159}
  (\bibinfo{publisher}{Academic Press}, \bibinfo{year}{2005}).
\newblock
  \urlprefix\url{http://www.sciencedirect.com/science/article/pii/S1049250X05520032}.

\bibitem{Jozsa2011}
\bibinfo{author}{Jozsa, R.} \& \bibinfo{author}{Linden, N.}
\newblock \bibinfo{title}{On the role of entanglement in quantum-computational
  speed-up}.
\newblock \emph{\bibinfo{journal}{Proceedings of the Royal Society of London A:
  Mathematical, Physical and Engineering Sciences}}
  \textbf{\bibinfo{volume}{459}}, \bibinfo{pages}{2011--2032}
  (\bibinfo{year}{2003}).

\bibitem{macchiavello}
\bibinfo{author}{Bru\ss{}, D.} \& \bibinfo{author}{Macchiavello, C.}
\newblock \bibinfo{title}{Multipartite entanglement in quantum algorithms}.
\newblock \emph{\bibinfo{journal}{Phys. Rev. A}} \textbf{\bibinfo{volume}{83}},
  \bibinfo{pages}{052313} (\bibinfo{year}{2011}).
\newblock \urlprefix\url{http://link.aps.org/doi/10.1103/PhysRevA.83.052313}.

\bibitem{beatrix4}
\bibinfo{author}{Gabriel, A.}, \bibinfo{author}{Huber, M.} \&
  \bibinfo{author}{Hiesmayr, B.~C.}
\newblock \bibinfo{title}{Criterion for k-separability in mixed multipartite
  systems}.
\newblock \emph{\bibinfo{journal}{Quantum Information and Computation (QIC) 10,
  No. 9 \& 10}} \bibinfo{pages}{829--836} (\bibinfo{year}{2010}).

\end{thebibliography}
\end{document}